\documentclass[aps,prc,twocolumn,superscriptaddress,showpacs]{revtex4}
\usepackage{graphicx}

\begin{document}

\title{Search for octupole correlations in $^{147}$Nd}
\author{E.~Ruchowska}
\email[]{Ewa.Ruchowska@ncbj.gov.pl}
\affiliation{National Center for Nuclear Research, Ho\.za 69, PL
00-681 Warsaw, Poland}
\author{H.~Mach}
\thanks{deceased}
\affiliation{National Center for Nuclear Research, Ho\.za 69, PL
00-681 Warsaw, Poland}
\author{M.~Kowal}
\affiliation{National Center for Nuclear Research, Ho\.za 69, PL
00-681 Warsaw, Poland}
\author{J.~Skalski}
\affiliation{National Center for Nuclear Research, Ho\.za 69, PL
00-681 Warsaw, Poland}
\author{W.A.~P\l\'ociennik}
\thanks{deceased}
\affiliation{National Center for Nuclear Research, Ho\.za 69, PL
00-681 Warsaw, Poland}
\author{B.~Fogelberg}
\affiliation{Department of Nuclear and Particle Physics, Uppsala
University, P.O. Box 535, S-75121 Uppsala, Sweden}
\date{\today}

\begin{abstract}
Properties of excited states in $^{147}$Nd have been studied with
multispectra and $\gamma \gamma$ coincidence measurements.
Twenty-four new $\gamma$-lines and three new levels have been
introduced into the level scheme of $^{147}$Nd. Lifetimes of eight
excited levels in $^{147}$Nd, populated in the $\beta$ decay of
$^{147}$Pr, have been measured using the advanced time-delayed
$\beta\gamma\gamma$(t) method. Reduced transition probabilities
have been determined for 30 $\gamma$-transitions in $^{147}$Nd.
Potential energy surfaces on the ($\beta_{2}$,$\beta_{3}$) plane
calculated for $^{147}$Nd using the Strutinsky method predict two
single quasiparticle configurations with nonzero octupole
deformation, with K=1/2 and K=5/2. We do not observe parity
doublet bands with K=5/2. For pair of opposite parity bands that
could form  the K=1/2 parity doublet we were able only to
determine lower limit of the dipole moment, $|D_0|\geq$0.02
e$\cdot fm$.
\end{abstract}
\maketitle

\section{\label{sec:Intro}Introduction}

Lowered excitation energies of the first 1$^{-}$ states, fast E1
transitions between the K$^\pi=0^{-}$ and ground state bands and
high $|D_0|$ values observed in the even-even $^{146-150}$Nd
isotopes constitute an evidence that these nuclei belong to the
octupole deformation region in lanthanides. Also theory assigns
these isotopes to the octupole region \cite{But91,But96}. This
same one should expect for the odd-N neodymium isotopes from this
mass region. In these isotopes one should observe parity doublet
bands connected by strong E1 transitions with high $|D_0|$
moments. However in Ref. \cite{Ruc10} in which the properties of
$^{149}$Nd isotope have been studied we have obtained very low
$|D_0|$ values for the lowest pair of opposite parity states which
should constitute parity doublet in this isotope. In the present
work we have undertaken investigations of octupole correlations in
$^{147}$Nd.

Excited  states in $^{147}$Nd have been previously studied in the
$\beta$-decay of $^{147}$Pr \cite{Dor75,Pin75,Shi93}, in the
neutron capture reaction \cite{Rou75,Rou76} and in the transfer
reactions \cite{Str77,Ham80,Lov80,Jas82}. Recently, high-spin
states in $^{147}$Nd have been studied with the use of the heavy
ion induced fusion-fission reaction \cite{Ven05}. Multipolarities
of several $\gamma$-transitions in $^{147}$Nd have been
established in Refs.~\cite{Rou76,Shi93} by internal-conversion
electron measurements. Lifetimes of the lowest levels in
$^{147}$Nd have been measured in Refs.~\cite{Ham80,Shi93}.
Experimental data concerning the $^{147}$Nd nucleus are gathered
in a recent compilation \cite{Nic09}. Despite of a rather big body
of experimental data on $^{147}$Nd no firm configuration
assignment for excited levels was proposed and there is no
experimental information on octupole correlations in $^{147}$Nd.
In order to obtain a better understanding of the low energy
structure and to get information on octupole strength in of
$^{147}$Nd we have measured lifetimes of the excited states in
this nucleus using the advanced time-delayed
$\beta\gamma\gamma(t)$ method \cite{Mac89,Mos89,Mac91}. The
experimental methods are briefly described in Section
\ref{sec:method}, while a new level scheme for $^{147}$Nd and the
lifetime results are presented in Section \ref{sec:results} of
this paper. In Section \ref{sec:discussion} experimental results
are discussed and results of theoretical calculations are
presented. In particular, results of potential energy calculations
on the ($\beta_{2}$,$\beta_{3}$) plane and theoretical values of
$|D_0|$ moments are shown.

\section{\label{sec:method}Experimental details}

Measurements were carried out at the OSIRIS on-line
fission-product mass separator at Studsvik in Sweden \cite{Fog92}.
Excited states in $^{147}$Nd were populated in the $\beta$-decay
of $^{147}$Pr, which was obtained via a chain of $\beta$-decays
starting from the $^{147}$Cs and $^{147}$Ba isotopes,

\[^{147}Cs \stackrel{0.230 s} {\rightarrow} \ ^{147}Ba \stackrel{0.894 s}
{\rightarrow} \ ^{147}La \stackrel{4.06 s}{\rightarrow} \ ^{147}Ce
\stackrel{56.4 s}{\rightarrow} \]
\[ {\rightarrow}\ ^{147}Pr \stackrel{13.4 min}{\rightarrow} \ ^{147}Nd.\]

\noindent The A=147 nuclei were produced in the fission reaction
of $^{235}$U induced by the thermal neutrons from the R2-0 reactor
at Studsvik. The $^{235}$U target consisted of about 1 g of
uranium dispersed in graphite. The A=147 activity, mass separated
from other fission products, was deposited onto an aluminized
Mylar foil in a moving-tape collection system at the center of the
experimental setup. Each measuring cycle was divided into eight
sequential time-bins, each lasting 40 s. To 'clean up' spectra
from the activities coming from the $^{147}$Pr predecessors our
radioactive samples were cumulated during first 135 s of each
cycle. Then the beam was deflected and the data were collected
during the rest of the cycle.

Two experiments have been performed. In the first one the
multispectra (MSS) and $\beta$-gated $\gamma \gamma$ coincidence
data have been collected. In this experiment one LEP Ge detector
with energy resolution FWHM of 0.6 keV at 81 keV, one 30\% HPGe
detector with FWHM of 2.1 keV and one 80\% HPGe detector with FWHM
of 2.9 keV at 1333 keV have been used. A 3 mm thick NE111A plastic
scintillator was used as a $\beta$-detector. About
1.2$\times$10$^{8}$ double coincident events have been collected.

In the second experiment lifetime measurements have been performed
using the advanced time-delayed $\beta \gamma \gamma $(t) method
\cite{Mac89,Mos89,Mac91}. In this method the fast timing
information was derived from coincidences between fast-response
$\beta$- and BaF$_2$ $\gamma$-detectors, while an additional
coincidence with a Ge $\gamma$-detector was used to select the
desired $\gamma$-cascade. In this experiment the setup consisted
of one BaF$_2$ detector, one HPGe detector with efficiency of 30\%
and FWHM of 2.1 keV at 1333 keV, and one $\beta$-detector. The
latter was a 3 mm thick $\Delta$E NE111A plastic detector to
ensure almost constant, independent of the $\beta$-particle
energy, time response of the fast timing system. About
2.2$\times$10$^{6}$ $\beta$-Ge-BaF$_2$(t) coincident events have
been collected. Several sets of the $\beta$-gated coincidence
$\gamma$-ray spectra from the Ge and BaF$_2$ detectors and triple
coincidence $\beta\gamma\gamma$(t) time-delayed spectra have been
sorted in the off-line analysis. Gating conditions set on the
$\beta$-spectrum were chosen to keep the time response constant in
the whole range of selected $\beta$ particle energies.

\begin{figure}
\includegraphics[width=9cm]{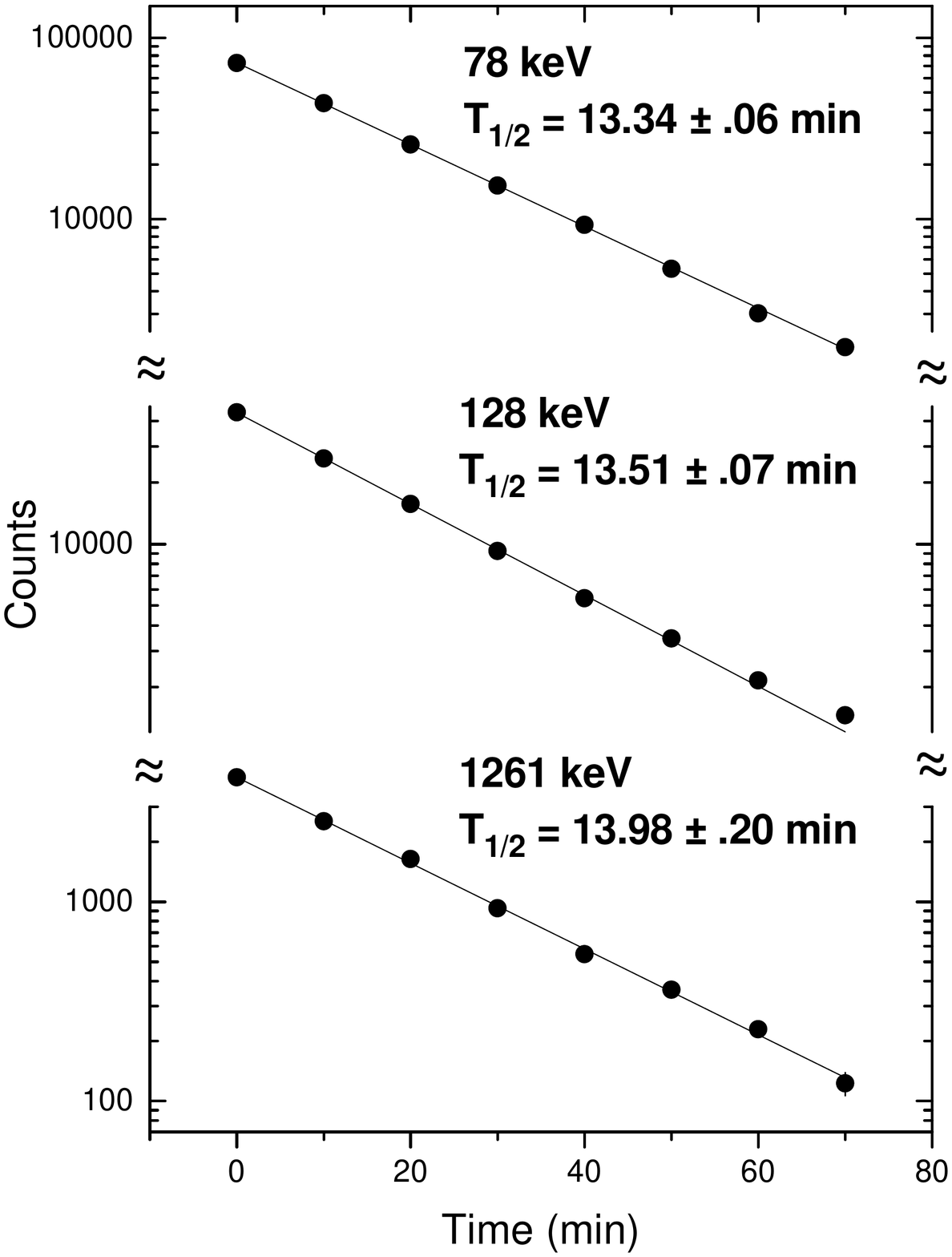}
\caption{\label{fig:mss} Examples of the decay curves of $\gamma$
rays from the $\beta$ decay of $^{147}$Pr obtained in the MSS
measurements.}
\end{figure}

\begin{figure*}
\includegraphics[angle=-90,width=14cm]{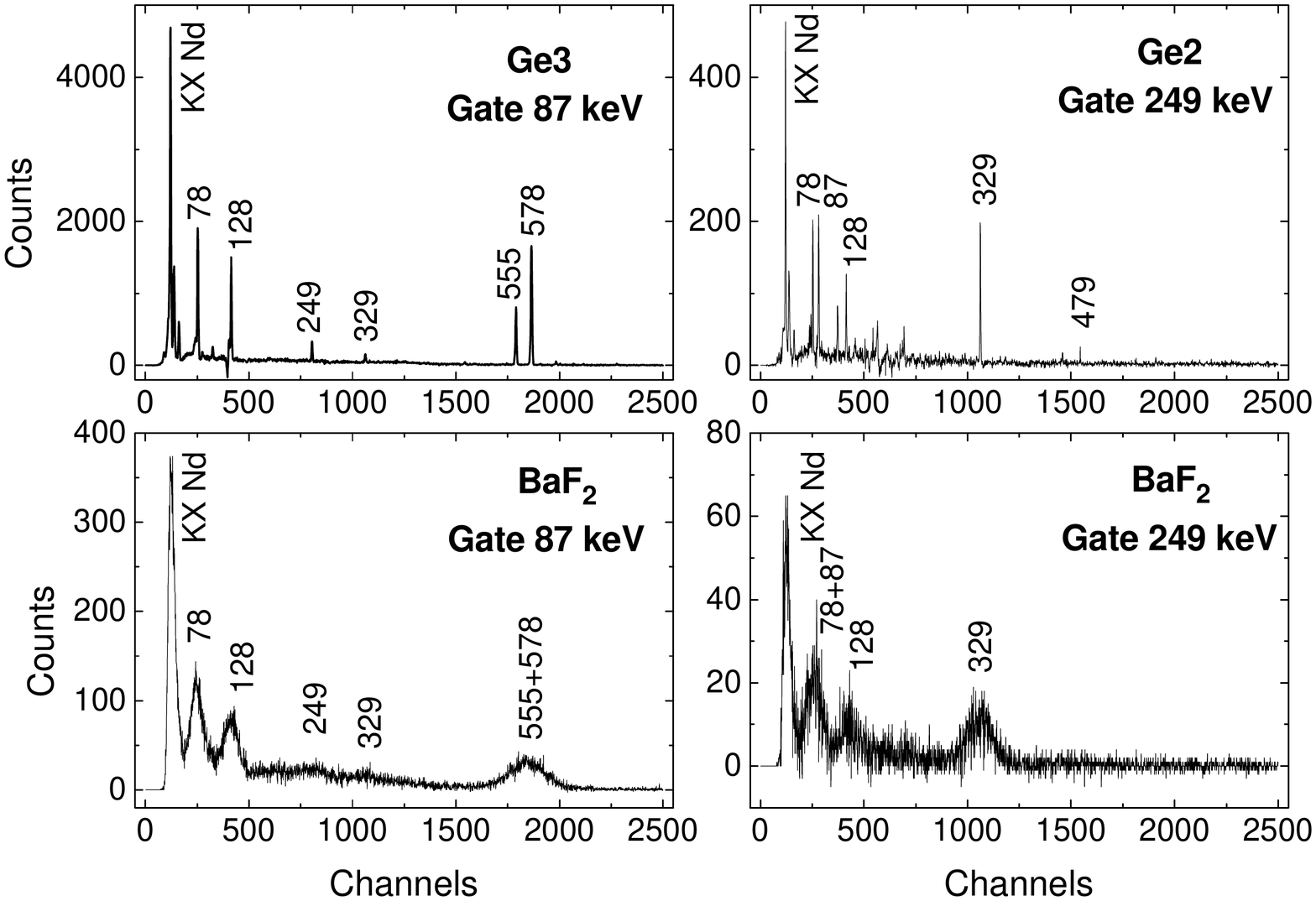}
\caption{\label{fig:gspe1} Coincident $\gamma$-ray spectra from
the Ge and BaF$_2$ detectors. Left top panel shows spectrum from
the 80\% HPGe detector gated by the 86.7 keV line in the LEP
detector. Right top panel shows spectrum from the 30\% HPGe
detector gated by the 249.0 keV line in the LEP detector. Panels
at the bottom show the corresponding coincident spectra sorted
onto the BaF$_2$ detector from the $\beta$-Ge-BaF$_2$ data.}
\end{figure*}

\begin{figure*}
\includegraphics[angle=-90,width=14cm]{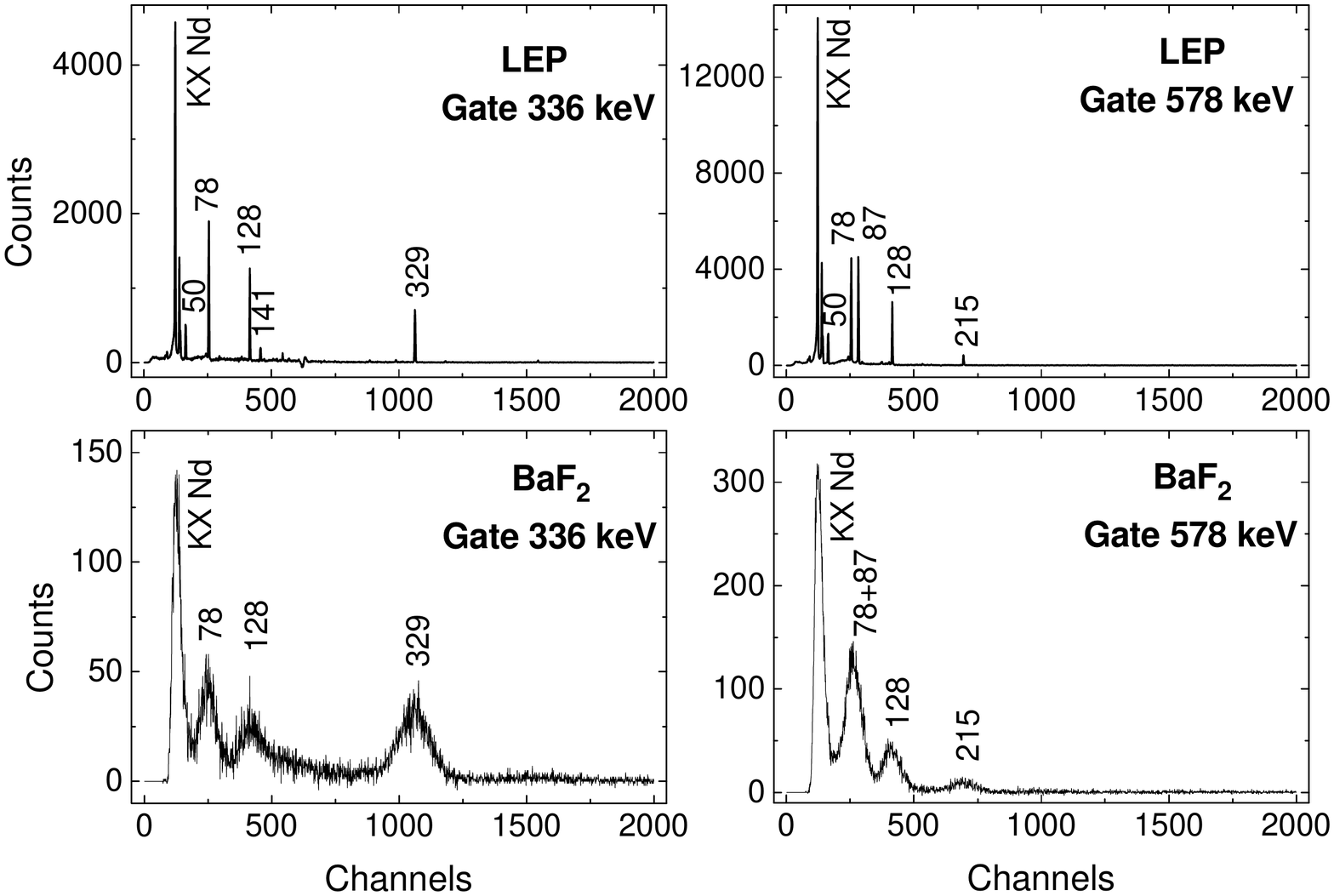}
\caption{\label{fig:gspe2} Coincident $\gamma$-ray spectra from
the Ge and BaF$_2$ detectors. Left top panel shows spectrum from
the LEP detector gated by the 335.7 keV line in the 30\% HPGe
detector. Right top panel shows spectrum from the LEP detector
gated by the 578.0 keV line in the 80\% HPGe detector. Panels at
the bottom show the corresponding coincident spectra from the
BaF$_2$ detector.}
\end{figure*}

\begin{table*}
\begin{footnotesize}
\caption{\label{tab:BML} Level lifetimes and experimental reduced
transition probabilities in $^{147}$Nd observed in the
$\beta$-decay of $^{147}$Pr. New $\gamma$ transitions are marked
with asterisk.} \vspace{.2cm}
\begin{ruledtabular}
\begin{tabular}{cccccccccc}
\noalign{\smallskip}Initial & I$_{i}^\pi K_{i}$ & T$_{1/2}$ &
T$_{1/2}$ & E$_\gamma$ & I$_\gamma$ & I$_{f}^\pi K_{f}$ &
$M\lambda$\footnotemark[1] &
$\alpha_{TOT}$\footnotemark[2] & B$_{exp}(M\lambda$)\footnotemark[3] \\
 level &  & this work & ref. \cite{Nic09} &  &  &  &  &  &  \\
$\lbrack$ keV $\rbrack$ &  & $\lbrack$ ps $\rbrack$ & $\lbrack$ ns
$\rbrack$ & $\lbrack$ keV $\rbrack$ &  &  &  &  &  \\
\noalign{\smallskip}\hline\noalign{\smallskip}
 49.9 & 7/2$^{-}$ 5/2 & 808(111)\ & 1.0(3) & 49.921(4) & 48(5) & 5/2$^{-}$ 5/2
 & M1 & 10.5 & $>$2.3$\cdot 10^{-2}$ \\
  &  &  &  &  &  &  & E2 & 31.3 & $<$2.3$\cdot 10^{4}$ \\
 127.9 & 5/2$^{-}$ 1/2 & 390(23) & 0.4(1) & 78.003(5) & 122(12) & 7/2$^{-}$ 5/2
 & M1 & 2.85 & $>$3.0$\cdot 10^{-2}$ \\
   &  &  &  &  &  &  & E2 & 5.4 & $<$1.6$\cdot 10^{4}$ \\
 &  &  &  & 127.941(6) & 100(10) & 5/2$^{-}$ 5/2 & M1 & 0.71 & $>$5.9$\cdot 10^{-3}$ \\
   &  &  &  &  &  &  & (E2) & 0.91 & $<$8.3$\cdot 10^{2}$ \\
 214.6 & 1/2$^{-}$ 1/2 & 4.59(22) ns & 4.53(6) & 86.698(5) & 69(7) & 5/2$^{-}$ 1/2
 & E2 & 3.9 & 4.8(5)$\cdot 10^{3}$ \\
 &  &  &  & 214.601(6) & 19(2) & 5/2$^{-}$ 5/2 & E2 & 0.155 & 1.4(2)$\cdot 10^{1}$ \\
 314.7 & 3/2$^{-}$ 3/2 & 54(7) & $\leq$0.1 & 100.090(8) & 5.5(5) & 1/2$^{-}$ 1/2
 & (M1) & 1.5 & 1.3(2)$\cdot 10^{-2}$ \\
 &  &  &  & 186.764(6) & 11.7(12) & 5/2$^{-}$ 1/2 & M1 & 0.25 & 4.4(7)$\cdot 10^{-3}$ \\
 &  &  &  & 264.74(4) & 4.0(4) & 7/2$^{-}$ 5/2 & E2 & .082 & 1.1(2)$\cdot 10^{2}$ \\
 &  &  &  & 314.728(6) & 255(25) & 5/2$^{-}$ 5/2 & M1 & 0.06 & 7.4($^{-41}_{+38}$)$\cdot 10^{-3}$ \\
   &  &  &  &  &  &  & E2 & 0.047 & 1.8($^{+7}_{-6}$)$\cdot 10^{3}$ \\
 463.6 & 3/2$^{-}$ 1/2 & 26(4) & $\leq$0.1 & 149.0(1) & 0.4(1) & 3/2$^{-}$ 3/2
 & M1 & 0.46 & 1.8(5)$\cdot 10^{-3}$ \\
 &  &  &  & 249.044(8) & 15.3(15) & 1/2$^{-}$ 1/2 & M1 & 0.112 & 8.8($^{-82}_{+57}$)$\cdot 10^{-3}$ \\
   &  &  &  &  &  &  & E2 & 0.095 & 1.3($^{+19}_{-13}$)$\cdot 10^{3}$ \\
 &  &  &  & 335.719(6) & 67(7) & 5/2$^{-}$ 1/2 & E2 & 0.039 & 3.1(6)$\cdot 10^{3}$ \\
   &  &  &  &  &  &  & (M1) & 0.052 & 1.9(11)$\cdot 10^{-3}$ \\
 &  &  &  & 413.675(9) & 13.3(13) & 7/2$^{-}$ 5/2 & E2 & 0.0208 & 2.3(4)$\cdot 10^{2}$ \\
 &  &  &  & 463.53(5) & 2.8(3) & 5/2$^{-}$ 5/2 & M1 & 0.022 & 4.1(8)$\cdot 10^{-4}$ \\
 604.5 &  1/2$^{-}$ 1/2 & 76(38) & $<$0.8 & 86.69(5)* & 1.6(2)* & 5/2$^{-}$ 3/2 & (E2) &
 3.85 & 1.0(5)$\cdot 10^{5}$ \\
 &  &  &  & 140.904(8) & 3.3(3) & 3/2$^{-}$ 1/2 & (M1) & 0.54 & 2.5(13)$\cdot 10^{-2}$ \\
 &  &  &  & 389.94(6)\footnotemark[4] & 0.6(2)\footnotemark[4] & 1/2$^{-}$ 1/2
 & (M1) & 0.0345 & 2.2(13)$\cdot 10^{-4}$ \\
 &  &  &  & 476.80(6)* & 1.6(2)* & 5/2$^{-}$ 1/2 & (E2) & 0.0138 & 2.0(10)$\cdot 10^{1}$ \\
 &  &  &  & 604.54(5) & 9.1(9) & 5/2$^{-}$ 5/2 & (E2) & 0.0075 & 3.5(18)$\cdot 10^{1}$ \\
 769.4 & 3/2$^{+}$ 1/2 & $\leq$16 &  & 305.70(4) & 2.4(2) & 3/2$^{-}$ 1/2 &
  (E1) & 0.0135 & $\geq7.5\cdot 10^{-6}$ \\
 &  &  &  & 454.64(5) & 1.4(1) & 3/2$^{-}$ 3/2 & (E1) & 0.005 & $\geq1.3\cdot 10^{-6}$ \\
 &  &  &  & 554.769(8) & 76(8) & 1/2$^{-}$ 1/2 & E1 & 0.00315 & $\geq4.0\cdot 10^{-5}$ \\
 &  &  &  & 641.392(8) & 215(22) & 5/2$^{-}$ 1/2 & E1 & 0.0023 & $\geq7.3\cdot 10^{-5}$ \\
 &  &  &  & 719.49(7) & 3.8(4) &  7/2$^{-}$ 5/2 & (M2) & 0.0205 & $\geq2.1\cdot 10^{2}$ \\
 &  &  &  & 769.32(8) & 4.8 (5) & 5/2$^{-}$ 5/2 & (E1) & 0.00155 & $\geq9.4\cdot 10^{-7}$ \\
 792.6 & 3/2$^{+}$  & $\leq$16 &  & 160.99(4) & 1.7(2) & 3/2$^{-}$ &
  (E1) & 0.072 & $\geq3.6\cdot 10^{-5}$ \\
 &  &  &  & 328.894(6) & 55(6) & 3/2$^{-}$ 1/2 & E)& 0.011 & $\geq1.4\cdot 10^{-4}$ \\
 &  &  &  & 477.841(8) & 56(7) & 3/2$^{-}$ 3/2 & E1 & 0.00435 & $\geq4.6\cdot 10^{-5}$ \\
 &  &  &  & 577.973(8) & 187(19) & 1/2$^{-}$ 1/2 & E1 & 0.00285 & $\geq8.7\cdot 10^{-5}$ \\
 &  &  &  & 664.60(7) & 3.1(3) &  5/2$^{-}$ 1/2 & (E1) & 0.0215 & $\geq9.5\cdot 10^{-7}$ \\
\end{tabular}
\end{ruledtabular}
\footnotetext[1] {$\gamma$-ray multipolarities and $\delta
^2$(E2/M1) mixing ratios are taken from Ref.~\cite{Nic09}.}
\footnotetext[2] {Total internal conversion coefficients are taken
from ref. \cite{Ban02}.} \footnotetext[3] {In units e$^2 fm^2$ for
E1 transitions, e$^2 fm^4$ for E2 transitions, $\mu^2_N$ for M1
transitions and $\mu^2_N fm^2$ for M2 transitions.}
\footnotetext[4] {$\gamma$-ray energy is from Ref.~\cite{Nic09}.
$\gamma$-ray intensity is re-calibrated from data of
Ref.~\cite{Nic09}.}

\end{footnotesize}
\end{table*}

\section{\label{sec:results}Experimental results}

\begin{figure*}
\includegraphics[width=15cm]{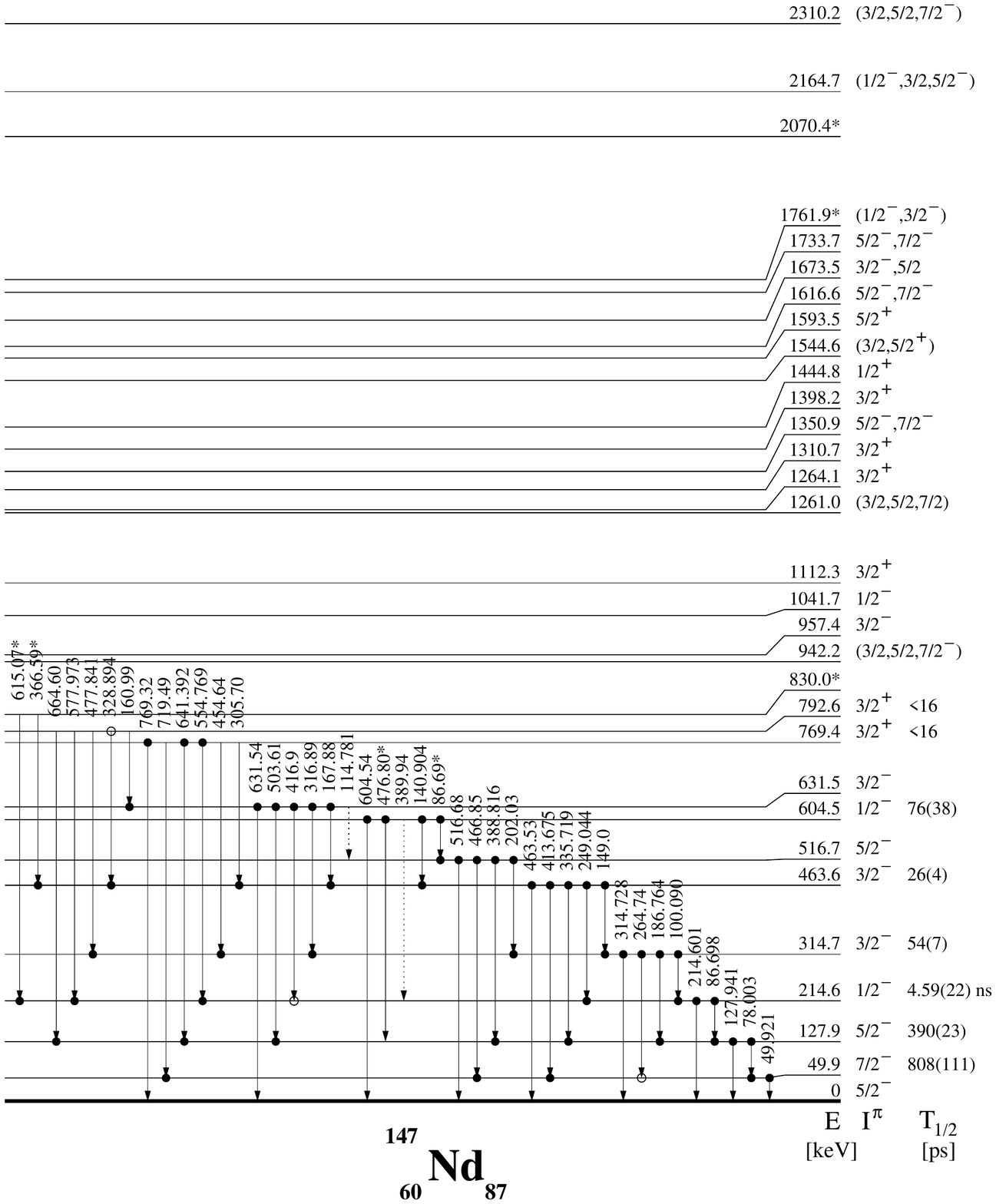}
\caption{\label{fig:schem1} The level scheme of $^{147}$Nd from
the $\beta$ decay of $^{147}$Pr. Full and empty circles indicate
strong and weak $\gamma\gamma$ coincidence relations,
respectively. Dotted lines mark transitions reported in Ref.
\cite{Nic09} but not observed in our data due to limited
statistics. Asterisks indicate new levels and transitions observed
in this work in the $\beta$ decay of $^{147}$Pr.}
\end{figure*}

The results for $^{147}$Nd are presented in Figs.~\ref{fig:mss}
$-$ \ref{fig:schem1} and in Tables \ref{tab:BML} and
\ref{tab:Rlin}. Gamma-lines were assigned to the $^{147}$Nd level
scheme based on the MSS  and $\gamma \gamma$ coincidence results.
Fig. \ref{fig:mss} presents examples of the decay curves for
$\gamma$-transitions emitted in the $\beta$-decay of $^{147}$Pr
obtained in the MSS measurements. The average half-life of
$^{147}$Pr obtained from these measurements, 13.39(4) min, is in
agreement with value 13.4(3) min given in Ref.~\cite{Nic09}.

The level scheme was constructed based on the $\beta$-Ge-Ge data.
Moreover several lines were placed in the level scheme based on
the energy fit. Examples of coincident $\gamma$-ray spectra
collected with the Ge detectors are shown in the upper panels of
Figs.~\ref{fig:gspe1} $-$ \ref{fig:gspe2}, while a new level
scheme for $^{147}$Nd is presented in Fig.~\ref{fig:schem1}. Level
and gamma-ray energies and gamma-ray intensities determined in the
present work are given Tables \ref{tab:BML} and \ref{tab:Rlin}.
Where available, the E2/M1 mixing ratios of $\gamma$-transitions
were taken from Ref.~\cite{Nic09}.

\begin{table*}
\begin{footnotesize}
\caption{\label{tab:Rlin} Energies and intensities of
$\gamma$-lines in $^{147}$Nd not listed in Table \ref{tab:BML}.
New $\gamma$ transitions and new levels are marked with
asterisks.}
\begin{ruledtabular}
\begin{tabular}{cccccc}
E$_\gamma$ & I$_\gamma$ & Initial level & E$_\gamma$ & I$_\gamma$ & Initial level\\
$\lbrack$ keV $\rbrack$ &  & $\lbrack$ keV $\rbrack$ & $\lbrack$
keV $\rbrack$ &  & $\lbrack$ keV $\rbrack$ \\
\hline\noalign{\smallskip}
 165.02(4)* & 1.5(2)* & 957.4 & 1036.38(8) & 1.9(2) & 1350.9 \\
 167.88(4) & 2.1(2) & 631.5 & 1046.06(8)* & 1.8(2)* & 1261.0 \\
 202.03(4) & 1.6(2) & 516.7 & 1080.4(1)* & 1.1(1)* & 1544.6 \\
 239.1(1)* & 0.6(1)* & 1350.9 & 1083.59(7) & 10.6(10) & 1398.2 \\
 310.7(1)* & 0.7(1)* & 942.2 & 1096.03(8) & 3.2(3) & 1310.7 \\
 316.89(5) & 1.2(1) & 631.5 & 1099.92(9) & 2.9(3) & 1616.6 \\
 366.59(5)* & 1.6(2)* & 830.0* & 1102.0(1)* & 1.3(1)* & 1733.7 \\
 388.816(8) & 18(2) & 516.7 & 1129.95(9) & 2.5(3) & 1593.5 \\
 416.9(1) & 0.9(1) & 631.5 & 1130.2(2)* & 0.6(1)* & 1761.9* \\
 437.25(5) & 1.6(2) & 1041.7 & 1136.53(7) & 19(2) & 1350.9 \\
 466.85(3) & 18(2) & 516.7 & 1152.6(1) & 2.3(2) & 1616.6 \\
 468.37(5)* & 1.9(2)* & 1261.0 & 1156.8(1) & 1.5(1) & 1673.5 \\
 478.51(8)* & 3.8(4)* & 942.2 & 1182.77(8) & 14.0(14) & 1310.7 \\
 491.70(5) & 4.8(5) & 1261.0 & 1214.3(1) & 4.2(4) & 1264.1 \\
 493.53(7) & 1.1(1) & 957.4 & 1217.1(1) & 2.0(2) & 1733.7 \\
 494.9(1)* & 0.7(1)* & 1264.1  & 1230.08(9) & 2.2(2) & 1544.6 \\
 503.61(5) & 4.8(5) & 631.5 & 1261.11(9) & 54(5) & 1261.0 \\
 516.68(3) & 15(1) & 516.7 & 1264.24(9) & 15.3(15) & 1264.1 \\
 518.43(8)* & 1.0(1)* & 1310.7 & 1298.5(1)* & 1.3(2)* & 1761.9* \\
 596.1(2) & 0.6(1) & 1112.3 & 1300.54(9) & 29(3) & 1350.9 \\
 615.07(6)* & 4.8(5)* & 830.0*  & 1302.0(1) & 2.9(3) & 1616.6 \\
 627.52(8) & 4.4(4) & 942.2 & 1310.64(9) & 7.1(7) & 1310.7 \\
 631.54(5) & 9.7(10) & 631.5 & 1358.9(1) & 2.6(3) & 1673.5 \\
 642.42(8) & 1.2(1) & 957.4 & 1398.4(2) & 1.4(3) & 1398.2 \\
 656.46(7) & 2.3(2) & 1261.0 & 1416.9(1) & 2.7(3) & 1544.6 \\
 706.12(7) & 6.9(7) & 1310.7 & 1465.2 (1) &1.6(2) & 1593.5 \\
 718.97(7) & 4.7(5) & 1350.9 & 1518.0(2)* & 1.6 (2)* & 2310.2 \\
 726.6(1) & 0.6(1) & 1041.7 & 1543.6(1) & 3.0(3) & 1593.5 \\
 746.45(7) & 3.6(4) & 1350.9 & 1547.2(1)* & 2.3(2)* & 1761.9* \\
 746.9(1)* & 1.3(1)* & 1264.1 & 1560.4(2) & 0.8(2) & 2164.7 \\
 793.93(5) & 17.8(1.8) & 1310.7 & 1593.7(1) & 2.5(3) & 1593.5 \\
 797.23(8) & 2.0(2) & 1112.3 & 1605.9(1) & 1.1(1) & 1733.7 \\
 800.49(8) & 2.3(2) & 1264.1 & 1616.6(2)* & 0.9(2)* & 1616.6 \\
 814.2(1) & 1.3(1) & 942.2 & 1623.5(1) & 3.4(3) & 1673.5 \\
 840.4(1) & 1.1(1) & 1444.8 & 1673.8(2) & 1.8(2) & 1673.5 \\
 847.02(8) & 3.0(3) & 1310.7 & 1683.7(2) & 1.6(3) & 1733.7 \\
 854.2(2) & 1.4(3) & 2164.1 & 1733.2(2) & 2.3(3) & 1733.7 \\
 881.51(8) & 4.6(5) & 1398.2 & 1755.6(2)* & 0.5(1)* & 2070.4* \\
 887.02(8) & 6.5(6) & 1350.9 & 1793.1(2) & 2.7(3) & 2310.2 \\
 934.45(9)* & 1.2(1)* & 1398.2 & 1846.4(3) & 0.6(1) & 2310.2 \\
 942.27(8) & 12.3(1.2) & 942.2 & 1942.5(2)* & 1.2(1)* & 2070.4* \\
 949.40(9) & 2.1(2) & 1264.1 & 1995.4(2) & 1.6(2) & 2310.2 \\
 957.8(2) & 1.3(1) & 957.4 & 2163.7(3) & 1.4(3) & 2164.7 \\
 981.0(1)* & 1.1(1)* & 1444.8 & 2310.3(4) & 0.9(2) & 2310.2 \\
 996.01(6) & 20(2) & 1310.7 &  &  & \\
\end{tabular}
\end{ruledtabular}
\end{footnotesize}
\end{table*}

In general our level scheme for $^{147}$Nd agrees well with the
one obtained by Shibata \textit{et al.}~\cite{Shi93}. Twenty-four
new $\gamma$-lines and three new levels have been introduced into
the decay scheme of $^{147}$Pr. They are marked with asterisks in
Figs.~\ref{fig:schem1}a$-$c and in Tables \ref{tab:BML} and
\ref{tab:Rlin}. New levels have energies 830.0, 1761.9 and 2070.4
keV. The latter two may correspond to the 1759 $\pm$ 5 and 2068
$\pm$ 5 keV levels, observed in the transfer reactions
\cite{Nic09}. Also the $^{147}$Pr $\beta$-decay data obtained with
the total absorption $\gamma$-ray spectrometer \cite{Gre97}
indicate existence of states with similar excitation energies in
$^{147}$Nd. For completeness weak transitions, which have not been
observed in our spectra due to limited statistics but which have
been reported in Ref.~\cite{Nic09}, are drawn with dotted lines in
the level scheme.

\begin{figure*}
\includegraphics[width=15cm]{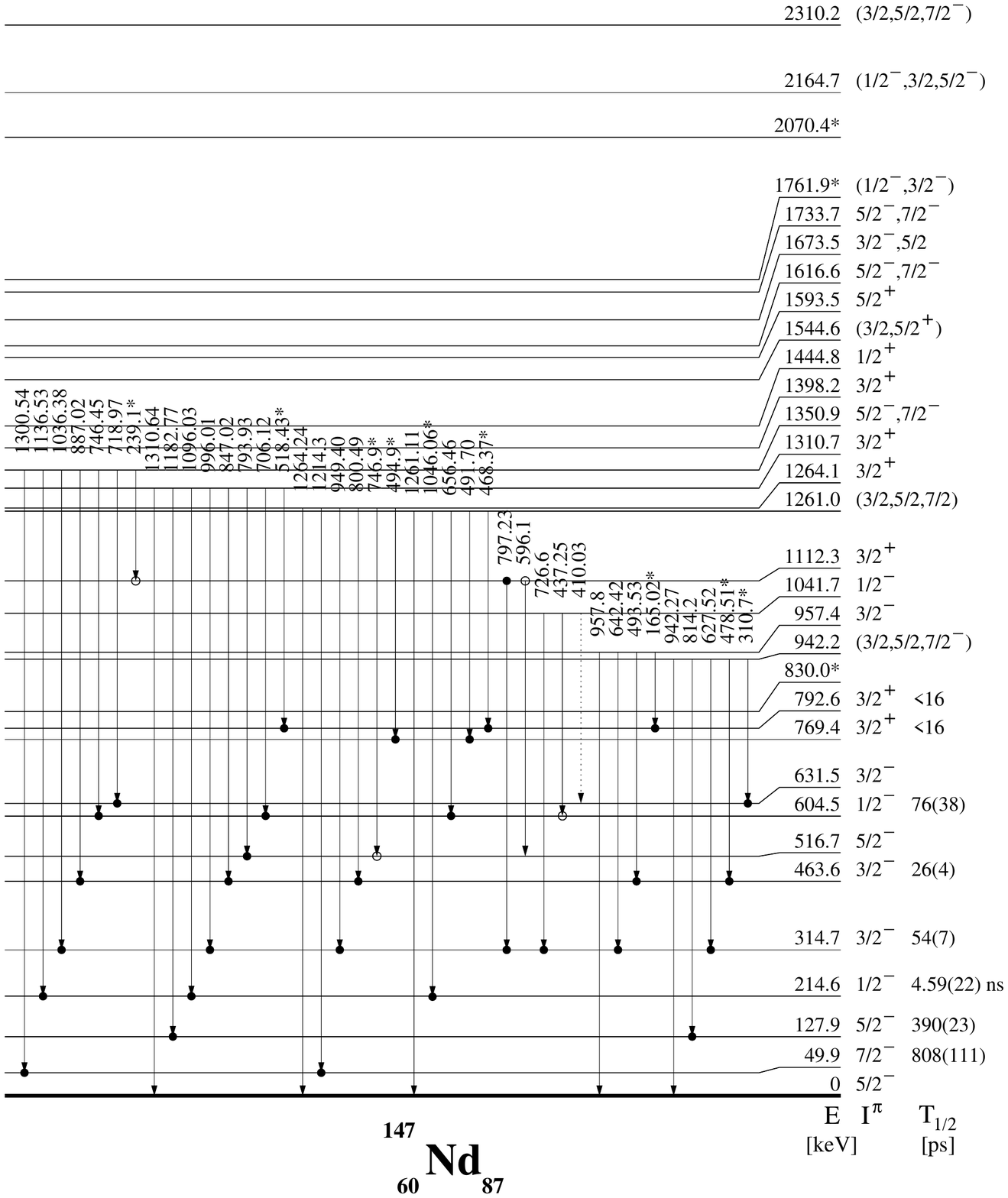}
\hspace{6cm} \textbf{Fig.~4:} \textit{(Continued).}
\end{figure*}

\begin{figure*}
\includegraphics[width=15cm]{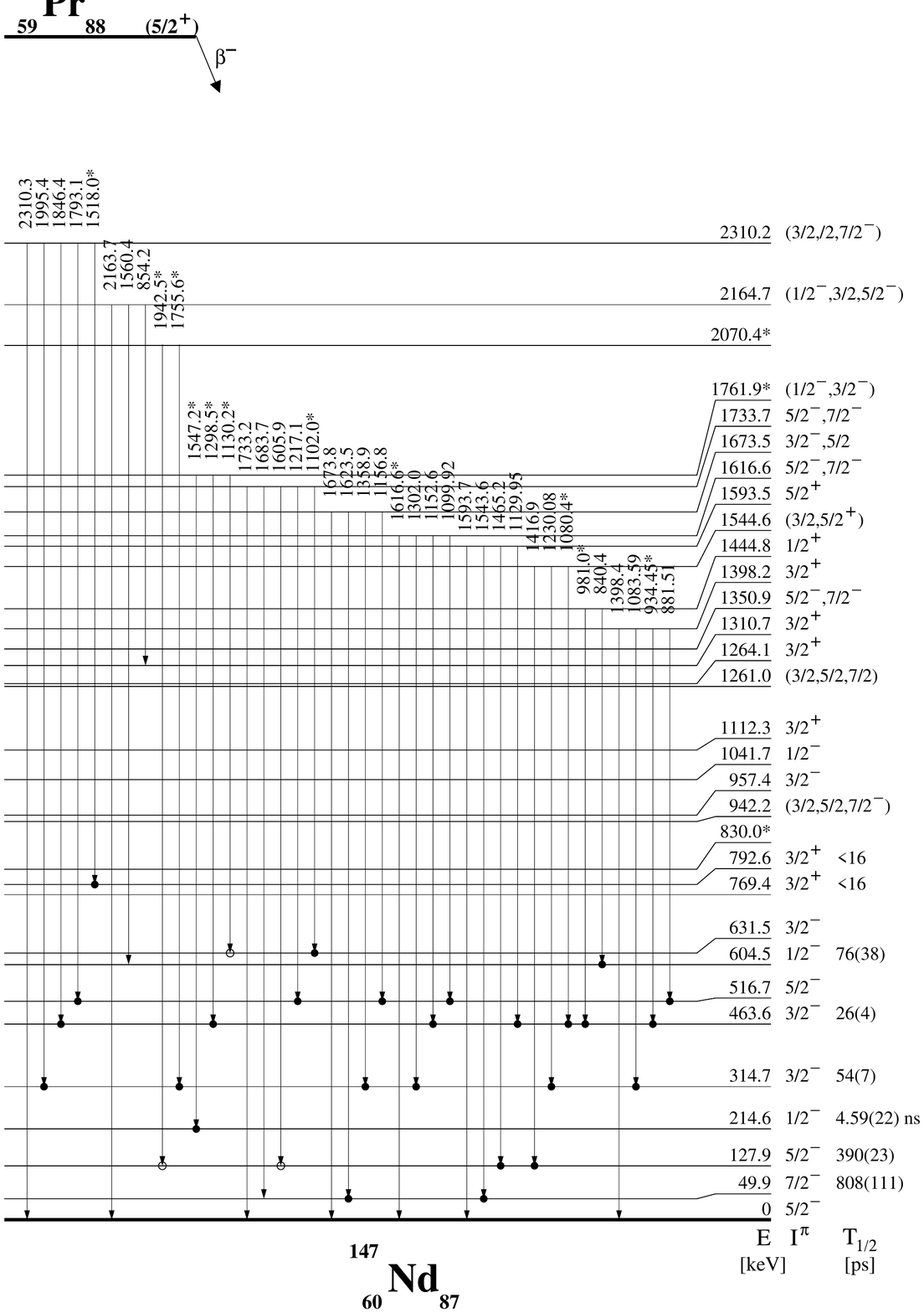}
\hspace{6cm} \textbf{Fig.~4:} \textit{(Continued).}
\end{figure*}

Level lifetimes were determined from the analysis of the triple
coincidence $\beta$-Ge-BaF$_2$ data. Examples of $\gamma$-ray
BaF$_2$ spectra coincident to the gating transitions selected in
the Ge detector are shown in the lower panels of
Figs.~\ref{fig:gspe1} $-$ \ref{fig:gspe2}.

\begin{figure}
\includegraphics[width=9cm]{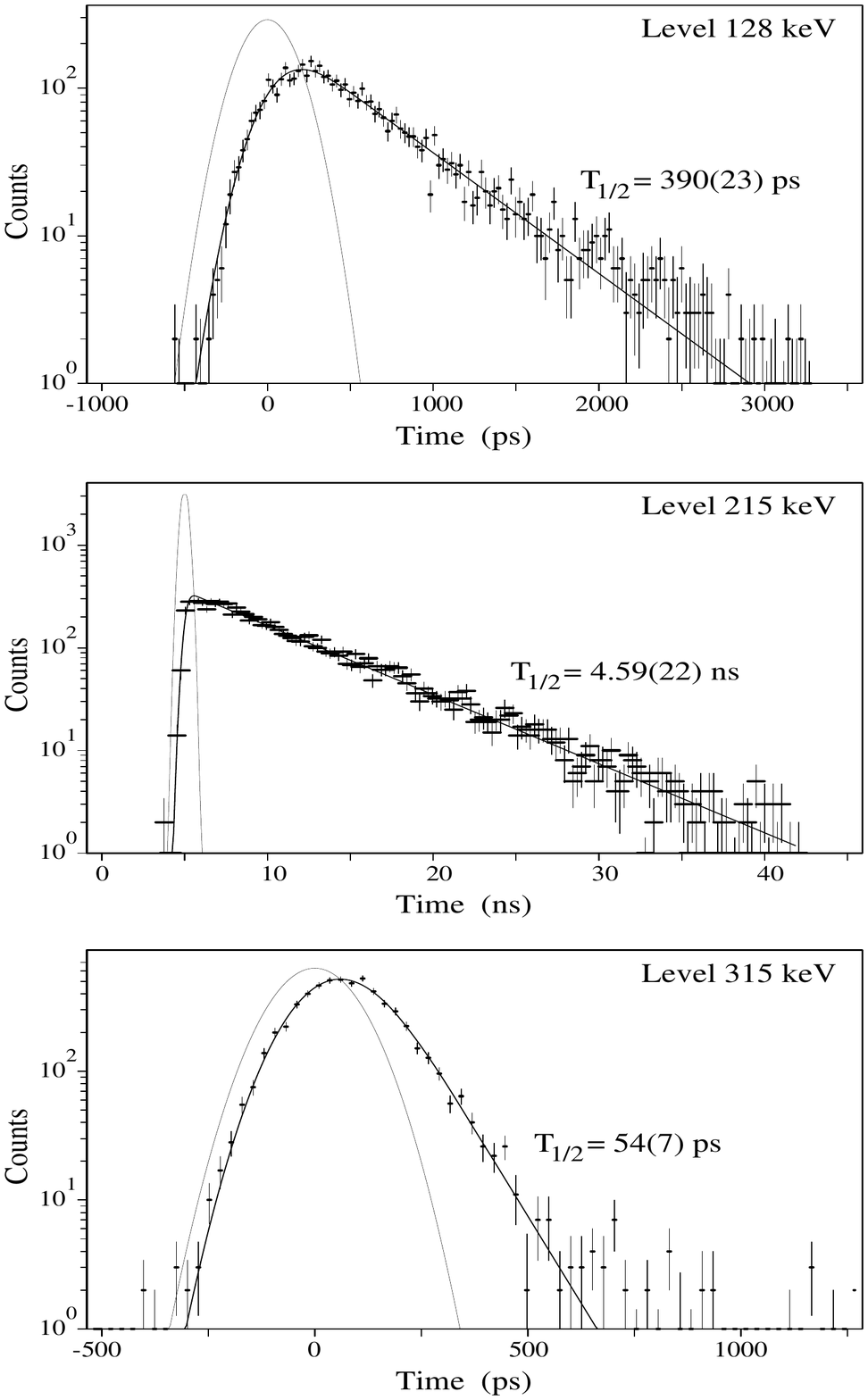}
\caption{\label{fig:tim1} Time-delayed $\beta\gamma\gamma$(t)
spectra showing slopes due to the lifetimes of the 127.9, 214.6
and 314.7 keV levels in $^{147}$Nd. Each figure shows experimental
points, prompt Gaussian spectrum and slope curve, which was fitted
in the de-convolution process.}
\end{figure}

Lifetimes longer than about 40 ps, which manifest themselves by a
strong asymmetry (or slope) on the delayed side of the time
spectra, have been determined using the de-convolution method
\cite{Mac89}. Examples of such time-delayed spectra for the 127.9,
214.6 and 314.7 keV levels in $^{147}$Nd are shown in
Fig.~\ref{fig:tim1}. The fitted function includes four free
parameters, namely position and Full-Width-at-Half-Maximum of
Gaussian, which approximates prompt time response of the timing
detectors, the half-life value and one parameter, which provides
an overall re-normalization between the experimental and fitted
time spectra. More details on the fitting procedures are given in
Ref.~\cite{Mac89}.  The half-lives for the levels at 49.9, 127.9,
214.6, 314.7 and 604.5 keV have been determined by the
de-convolution method.

\begin{figure}
\includegraphics[angle=-90,width=8cm]{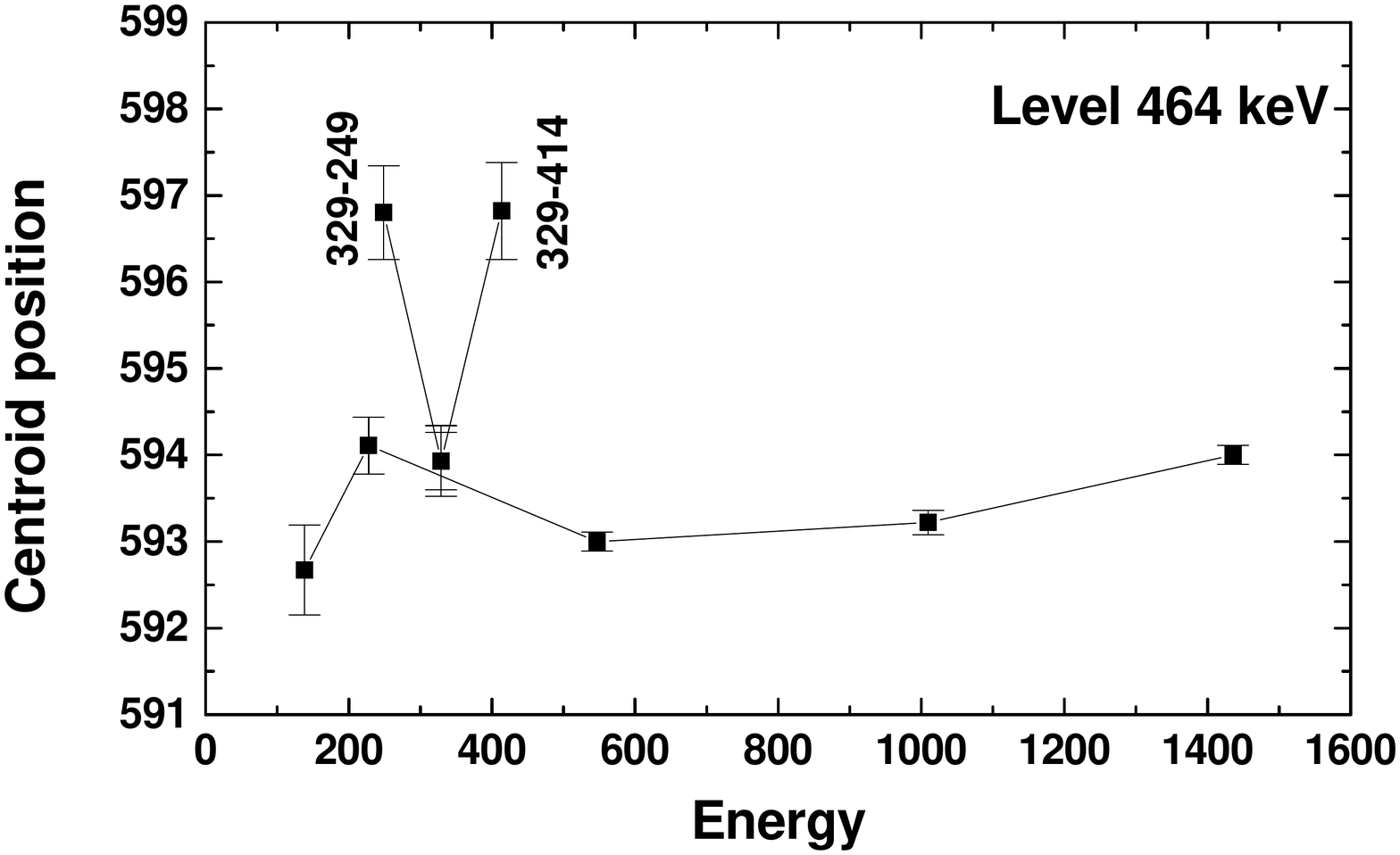}
\caption{\label{fig:tim2} Determination of the half-life for the
463.6 keV level in $^{147}$Nd by centroid shift technique. The
reference time curve made for $^{138}$Cs calibration source is
shown together with the centroid positions of time-delayed spectra
marked by the energies (in keV) of $\gamma$-ray gates in the Ge
and BaF$_2$ detectors which were used to generate them. The shift
of the time-delayed centroid from the reference curve gives the
mean-life of the level (calibration 12.85 ps/ch); see text for
details.}
\end{figure}

Lifetimes shorter than 40 ps were measured using the centroid
shift method \cite{Mac89}, in which the mean lifetime
$\tau$=T$_{1/2}$/ln$2$ is determined as a shift of the centroid of
the time-delayed spectrum from the prompt curve at a given
$\gamma$-ray energy. This method is based on the concept of a two
$\gamma$-ray cascade \cite{Mac89}. When a gate on the Ge detector
is set on the bottom member of the $\gamma$-ray cascade,
de-exciting intermediate state, and a gate on the BaF$_2$ detector
is set on the top $\gamma$-ray, de-exciting high energy excited
state in a nucleus, one obtains first centroid position of the
time-delayed $\beta$-BaF$_2$ spectrum giving a reference point at
the energy of the top $\gamma$-ray. By reversing gates on the
BaF$_2$ and Ge detectors, namely by selecting the top member of
the cascade in the Ge detector, and the bottom member in the
BaF$_2$ crystal we determine second time centroid position at the
energy of the bottom $\gamma$-ray. By shifting both centroids to
place the first centroid on the reference time curve we obtain the
meanlife of the intermediate level as a distance of the second
centroid from the reference curve. The reference time curve was
constructed from the $\beta$-gated $\gamma \gamma$ data measured
for the $^{138}$Cs calibration source in the same experimental
conditions as our data. This way a half-life of 26(4) ps was
obtained for the 463.6 keV level as an average from the time
spectra obtained with the 328.9-249.0, 328.9-335.7, 328.9-413.7
and 887.0-335.7 keV Ge-BaF$_2$ gate combinations. Examples of
short half-life determination using reference curve are given in
Fig.~\ref{fig:tim2}. Besides we have obtained half-life limits for
the 769.4 and 792.6 keV levels in $^{147}$Nd.

The half-life values obtained in present work for excited states
in $^{147}$Nd are given on the level scheme and are listed in the
third column in Table \ref{tab:BML}. Up to now only half-lives for
first three excited states had been determined in $^{147}$Nd
\cite{Nic09}. Our half-lives obtained for these levels agree very
well with  values given in Ref. \cite{Nic09}. For next 3 excited
states only the nanosecond half-life limits had been previously
known (see column 4 of Table \ref{tab:BML}). We have determined
much more precise half-life values for them.

\begin{figure}
\includegraphics[width=8cm]{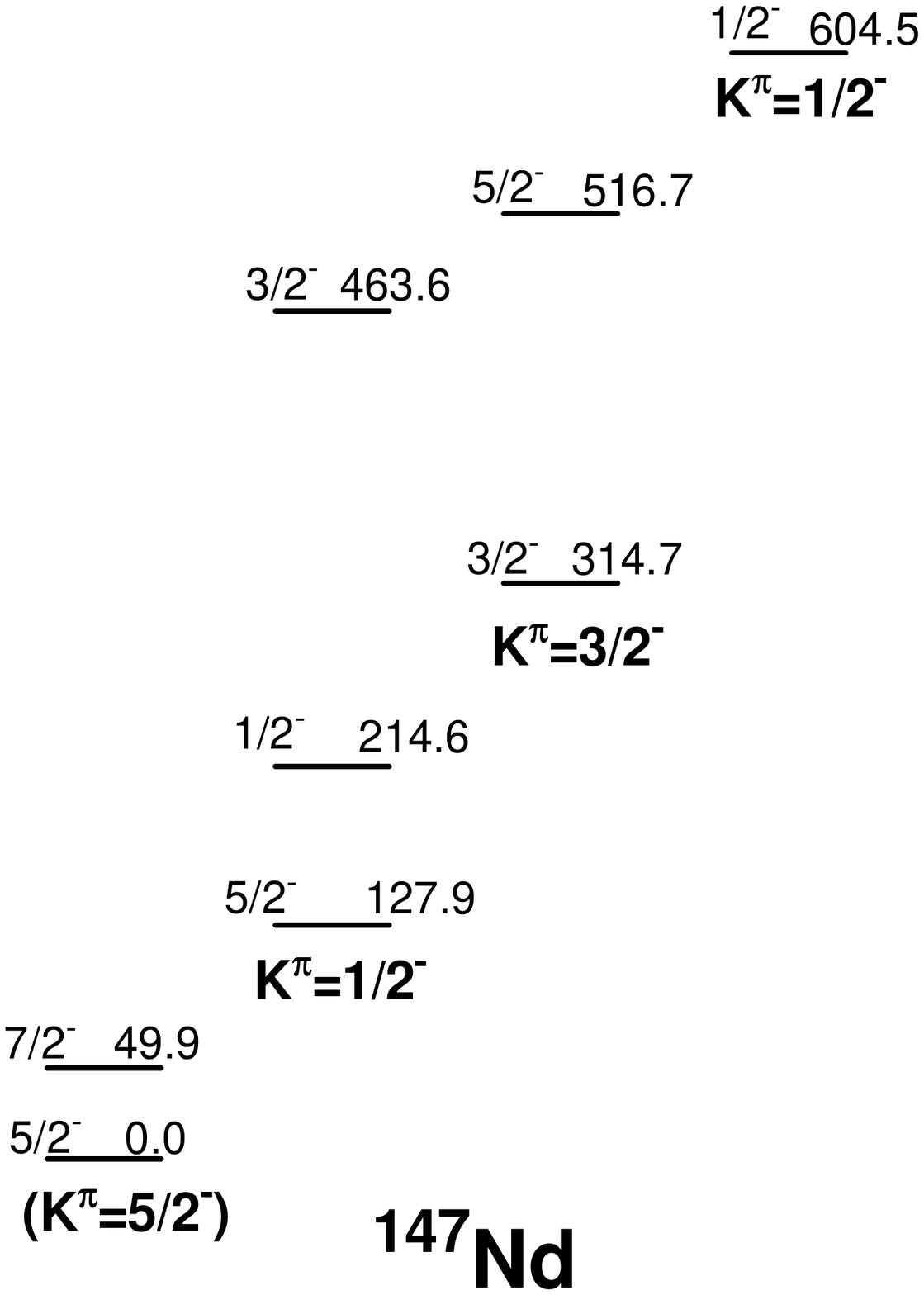}
\caption{\label{fig:levels} Interpretation of the band structure
in $^{147}$Nd.}
\end{figure}

Reduced transition probabilities have been determined from the
level lifetimes and the relative $\gamma$-ray intensities using
total internal conversion coefficients from ref. \cite{Ban02}.
Values of reduced transition probabilities obtained in this work
for 30 transitions in $^{147}$Nd are listed in the last column of
Table \ref{tab:BML}.

\begin{figure*}
\includegraphics[width=18cm]{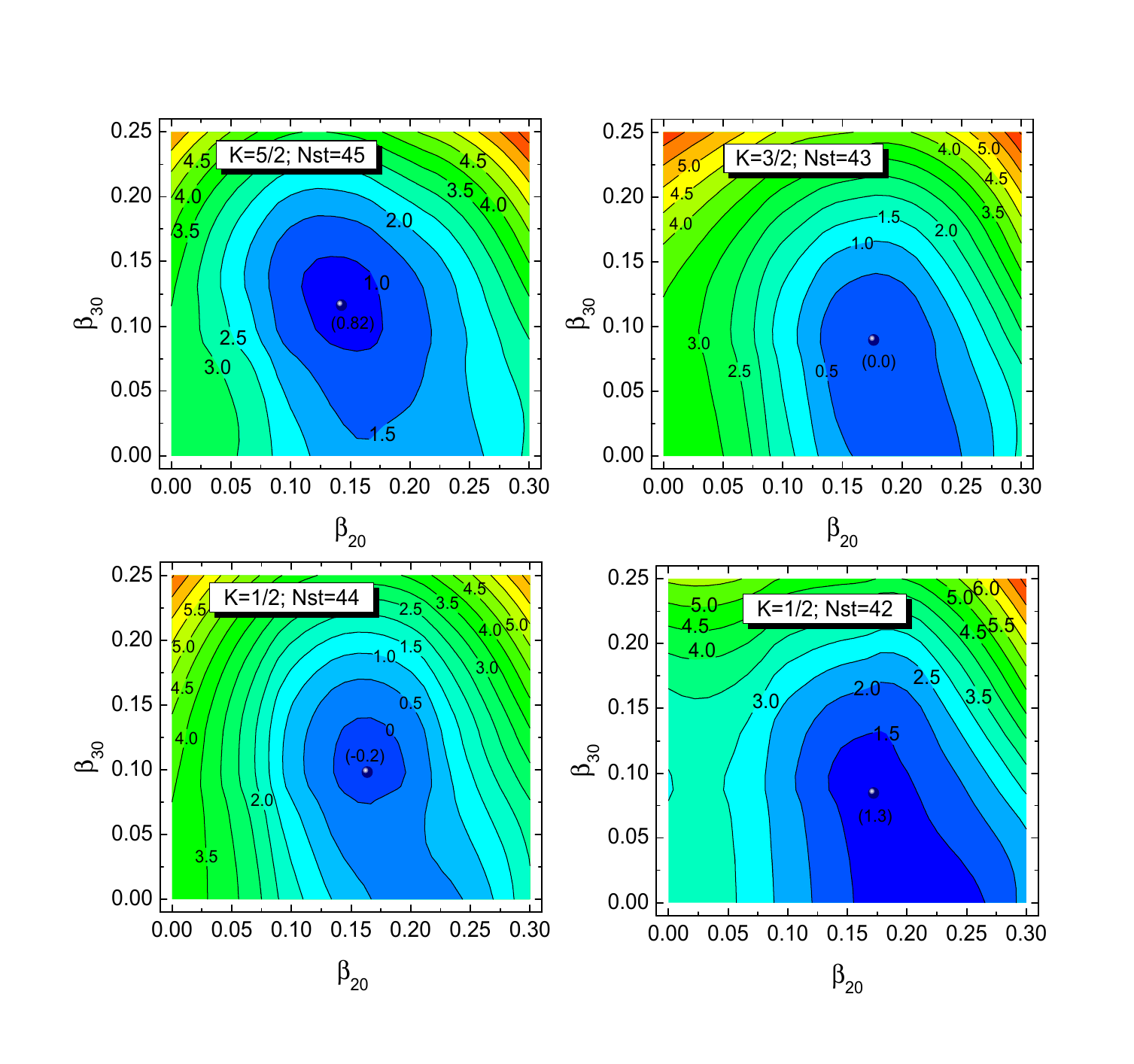}
\caption{\label{fig:maps} Potential energy surfaces calculated for
the lowest single quasi-particle configurations in $^{147}$Nd with
K = 1/2, 3/2 and 5/2 and single particle state numbers Nst = 42,
43 , 44 and 45. The energy distance between the contour lines is
equal to 0.5 MeV.}
\end{figure*}

\section{\label{sec:discussion}Discussion}

\subsection{\label{sec:configurations}Band structure in $^{147}$Nd}

The occurrence of strong octupole correlations in odd-A nuclei is
manifested by the presence of parity doublet bands connected by
the enhanced E1 transitions with the reduced transition
probabilities B(E1) of order $10^{-3}$ $e^2fm^2$ \cite{But96}. In
the case of parity doublet bands with K$^\pi$=1/2$^{\pm}$ the
decoupling parameters are expected to be of equal magnitudes and
of opposite signs.

Our potential energy calculations from Sec.~\ref{sec:surfaces}
performed for $^{147}$Nd predict two octupole-soft states with
K=1/2 and K=3/2 and two otupole-deformed states with K=1/2 and
K=5/2 at lowest excitation energies. So we should expect in our
data two parity doublet bands, one with K$^\pi$=1/2$^{\pm}$ and
one with K$^\pi$=5/2$^{\pm}$.

Ground state and the 49.9 keV state with spin 7/2$^{-}$ are
candidates for the K$^\pi$=5/2$^{-}$ band per analogy with
$^{149}$Nd \cite{Ruc10}. Magnetic $|g_{K}-g_{R}|$ factor for this
band deduced from the B(M1) value for 49.9 keV transition treated
as pure M1 would be 0.33(2), compared to 0.40(2), obtained in
$^{149}$Nd.

The B(E1) branching ratios for 554.8 and 305.7, 641.4 and 305.7
and 641.4 and 544.8 keV transitions deexciting the 769.4 keV level
to the 214.6, 463.6 and 127.9 keV states are equal 5.3(7), 9.7(13)
and 1.8(3), respectively. They agree very well with the
theoretical branching ratios 5.0, 9.0 and 1.8, calculated with
Alaga rule \cite{Ala55} assuming K=1/2 for all four states. We
suggest that 214.6, 463.6 and 127.9 keV states form the
K$^\pi$=1/2$^{-}$ band in $^{147}$Nd with $E_0$ parameter,
rotational parameter A and decoupling parameter {\sl a} equal
285.7 keV, 7.93 keV and 9.47, respectively. The quadrupole moment
$Q_0$ for this band, calculated from from the reduced transition
probability for 86.7 keV line given in Table \ref{tab:BML}, is
equal 284(16) $e\cdot{fm}^2$.

The B(M1) branching ratio for 186.8 and 100.1 keV transitions (see
Table \ref{tab:BML}) de-exciting the 314.7 keV level to the 5/2
and 1/2 states of K$^\pi$=1/2$^{-}$ band is equal 0.33(8). It
agrees with theoretical value 0.20 obtained with K=3/2 taken for
the 314.7 keV level. The B(E1) branching ratio for 797.2 and 596.1
keV transitions de-exciting the 1112.3 keV level to the 314.7 and
516.7 keV states is equal 1.4(3). It agrees with theoretical value
1.5 if all three states have K=3/2. Similar situation is obtained
for for 1083.6 and 881.5 keV transitions de-exciting the 1398.2
keV level where we obtain 1.2(2) and 1.5 for experimental and
theoretical values, respectively. Next B(E1) branching ratio for
the 949.4 and 746.9 keV transitions de-exciting the 1264.1 keV
level to the 314.7 and 516.7 keV states, equal 0.8(1), agrees with
theoretical value 0.7 when the 1264.1 keV level has
K$^\pi$=1/2$^{+}$ and the 314.7 and 516.7 keV states have
K$^\pi$=3/2$^{-}$. Above branching ratios indicate that the 314.7
and 516.7 keV states constitute a band with K$^\pi$=3/2$^{-}$.
Proposed band structure for $^{147}$Nd is shown in
Fig.~\ref{fig:levels}.

\subsection{\label{sec:octcorrel}Octupole correlations in $^{147}$Nd}

We do not observe the K$^\pi$=5/2$^+$ band at the lowest
excitation energies in $^{147}$Nd, although its population should
be enhanced due to the 5/2$^{+}$ ground state spin of the mother
nucleus. Neither we see the K$^\pi$=1/2$^{+}$ band close to the
$^{147}$Nd ground state. The lowest excited states with opposite
parity that could be candidates for the parity doublet members are
the I$^\pi$K=3/2$^+$1/2 769.4 keV and the I$^\pi$=3/2$^+$ 792.6
keV levels. We were able to determine only the upper limits of
their half-lives, T$_{1/2}\leq$16 ps, for both of them. The lower
limits of the B(E1) values for the $\gamma$-transitions
depopulating these two levels are given in Table \ref{tab:BML}.

A generally adopted way of comparison of the E1 strength over a
wider range of nuclei is offered by the intrinsic electric dipole
moment, $|D_0|$, which removes the spin dependence affecting the
B(E1) rates. Assuming a strong-coupling limit and an axial shape
of the nucleus, the electric dipole moment, $|D_0|$, is defined
(for K$\neq 1/2$) via the rotational formula:

\begin{equation}
\label{BE1}
 B(E1)={\frac{3}{4\pi}}D^{2}_{0}\langle
I_{i}K10|I_{f}K\rangle ^{2}.
\end{equation}

\noindent The $|D_0|$ moment is used as a convenient parameter for
the inter-comparison even for transitional nuclei.

The lower limit of $|D_0|$ obtained for transitions de-exciting
the level at 769.4 keV in $^{147}$Nd to the 463.6, 214.6 and 127.9
keV states from the lower K$^\pi$=1/2$^{-}$ band is
$|D_0|\geq$0.02 $efm$. In case of the 792.6 keV level, if this
level has K=1/2 the lower limit for the 328.9, 578.0 and 664.6 keV
transitions de-exciting this level to the lower K$^\pi$=1/2$^{-}$
band is $|D_0|\geq$0.04 $efm$ and if it has K=3/2 it is
$|D_0|\geq$0.02 $efm$ for the 477.8 keV transition to the
K$^\pi$=3/2$^{-}$ band.

Experimental information on $|D_0|$ moments in the odd-N isotopes
in lanthanides is very scarce. Reported $|D_0|$ values are very
low, much lower than in the even-even neighbors, or no parity
doublet bands are found at low excitation energies. Low $|D_0|$
values have been determined for $^{149}$Nd \cite{Ruc10}(see also
Table \ref{tab:D0}) and for $^{149}$Ce~\cite{Syn03}. Moreover, no
candidates for parity doublets have been found at the lowest
excitation energies in the odd N barium isotopes,
$^{145}$Ba~\cite{Rza12} and $^{147}$Ba~\cite{Rza13}, lying at the
very center of the octupole region in lanthanides. Even the ground
state spins of these two isotopes can be reproduced without
involving octupole correlations~\cite{Rza12,Rza13}. In $^{147}$Nd
we were only able to determine lower limit of $|D_0|$ for the
K=1/2 parity doublet band, however non-observation of the K=5/2
parity doublet band may suggest that the $|D_0|$ values in this
isotope are as low as in $^{149}$Nd and $^{149}$Ce.

The $|D_0|$ behavior in the odd-N isotopes in lanthanides is
different from smooth dependence versus neutron number N predicted
by the theory (see Table \ref{tab:D0} for Nd isotopes) or observed
in thorium isotopes in the actinide octupole region (see Fig. 10
in Ref.~\cite{Ruc10}). Low $|D_0|$ values in the odd-N isotopes do
not result from quenching caused by opposite signs of macroscopic
and shell-correction components of the dipole moment (see
Sec.~\ref{sec:D0}).

\begin{table*}
\begin{footnotesize}
\caption{\label{tab:D0} Experimental and theoretical $|D_0|$
values in Nd isotopes (in $efm$).}
\begin{ruledtabular}
\begin{tabular}{cccccccc}
N & 84 & 85 & 86 & 87 & 88 & 89 & 90 \\
Isotope & $^{144}$Nd & $^{145}$Nd & $^{146}$Nd & $^{147}$Nd &
$^{148}$Nd &
$^{149}$Nd & $^{150}$Nd \\
\hline\noalign{\smallskip} Experiment & 0.12(1)\footnotemark[1] &
& 0.16(4)\footnotemark[2] & $\geq$0.02\footnotemark[3] &
0.21(4)\footnotemark[2] & 0.06(2)\footnotemark[4]
& 0.26(5)\footnotemark[2] \\
Theory\footnotemark[3] &  & 0.15 & 0.17 & 0.26 & 0.30 & 0.36 & \\
\end{tabular}
\end{ruledtabular}
\footnotetext[1] { determined from level lifetime data given in
the www.nndc.bnl.gov data base.} \footnotetext[2] { from
Ref.~\cite{But96}.} \footnotetext[3] { this work.}
\footnotetext[4] { from Ref.~\cite{Ruc10}.}
\end{footnotesize}
\end{table*}

Recently new high spin data have been reported for the $^{147}$Ce
nucleus in Ref.~\cite{JLi14}. The $|D_0|$ moments calculated from
the B(E1)/B(E2) values given in this reference are 0.18(1) and
0.21(2) $efm$ for the 31/2$^{-}$, 2703.1 keV and 35/2$^{-}$,
3264.0 keV levels, respectively. These values are comparable to
the average $|D_0|$ values of 0.19(3) and 0.21(2) $efm$ obtained
from the B(E1)/B(E2) values in  the neighboring
$^{146}$Ce~\cite{Pek97} and $^{148}$Ce~\cite{Nic14} isotopes,
respectively.

It seems that considered opposite parity states in $^{147,149}$Nd
and in $^{149}$Ce may not constitute parity doublets and the main
octupole strength is located at higher excitation energies. It is
possible that the presence of odd neutron blocks octupole
correlations at low excitation energies in the odd N isotopes.
More experimental information on the parity doublets and dipole
moments in $^{147}$Nd and in other odd N isotopes in lanthanides
is needed to understand the $|D_0|$ moments behavior in nuclei
from this region.

\subsection{\label{sec:surfaces}Potential energy surfaces}

Energy surfaces over ($\beta_{20}$,$\beta_{30}$) plane were
calculated for the lowest single-particle configurations in
$^{147}$Nd by using the macroscopic-microscopic method.
Macroscopic energy was calculated using the Yukawa plus
exponential model \cite{Kra79} with parameters specified in
\cite{Mun01}. A deformed Woods-Saxon potential potential
\cite{Cwi87} with the universal set of parameters \cite{Dud81} was
used to calculate the microscopic energy.

We used a $\beta$ parametrization, $\beta$ =
($\beta_{20},\beta_{30},\beta_{40},\beta_{50},$
$\beta_{60},\beta_{70},\beta_{80}$), in which the shape of a
nucleus is described by the formula:
\begin{equation}
\label{shape}
 R(\theta) = R_0 c(\{\beta\}) \{ 1 +
 \sum_{\lambda=2}^8 \beta_{\lambda 0} {\rm Y}_{\lambda 0}(\theta) \} .
\end{equation}
where $R(\theta)$ is the nuclear radius and $c(\{\beta\})$ is the
volume-fixing factor.

Considered single particle configurations were chosen by blocking
the odd neutron on a desired single particle state with a given
$K$. Energies were calculated on a grid of 1361367 points defined
by deformation values:
\begin{center}
\begin{eqnarray}
\beta_{20} &=&   0.00 \ (0.05) \ 0.30 ;\nonumber\\
\beta_{30} &=&   0.00 \ (0.05) \ 0.30 ;\nonumber\\
\beta_{40} &=&  -0.20 \ (0.05) \ 0.20 ;\nonumber\\
\beta_{50} &=&  -0.20 \ (0.05) \ 0.20 ;\nonumber\\
\beta_{60} &=&  -0.15 \ (0.05) \ 0.15 ;\nonumber\\
\beta_{70} &=&  -0.15 \ (0.05) \ 0.15 ;\nonumber\\
\beta_{80} &=&  -0.15 \ (0.05) \ 0.15 ;
\end{eqnarray}
\end{center}
where the step in each deformation is given in parenthesis. For
each pair ($\beta_{20},\beta_{30}$) energy was then minimized with
respect to $\beta_{40}$ $-$ $\beta_{80}$. Energy surfaces obtained
for four single-particle configurations in $^{147}$Nd are
presented in Fig.~\ref{fig:maps}. They show equilibrium
deformations: $\beta_{20}$ of 0.13 - 0.18 and $\beta_{30}\approx
0.10$. Two configurations, $K=1/2$ and 3/2 (energy levels no. 42
and 43) with shallow octupole minima may be called octupole-soft.
Slightly more pronounced, $\approx 0.5$ MeV deep, octupole minima
were obtained for configurations with $K=1/2$ and 5/2 (levels no
44 and 45; the Fermi level in $^{147}$Nd lies in-between levels no
43 and 44).

Energy calculations for two even-A neighbors of $^{147}$Nd show
octupole minima for both $^{146}$Nd and $^{148}$Nd ground states.
Thus, our calculations suggest that $^{147}$Nd lies inside the
lanthanide octupole collective region.

\subsection{\label{sec:D0}Theoretical electric dipole moments}

Reduced probabilities of electromagnetic (EM) transitions between
the rotational band built on the one-phonon state and the g.s.
band can be calculated assuming the fixed structure of both the
phonon and the collective rotor \cite{Boh75}. For an operator
${\cal M}$ of the multipolarity $\lambda$ one has in odd-A nucleus
\begin{eqnarray}
\label{Btran}
  B(\lambda;K_1,I_1\rightarrow K_2,I_2)=2\langle I_1 K_1 \lambda K_2-K_1\mid
I_2 K_2\rangle^2 & & \\ \nonumber
  \mid\langle K_2\mid{\cal M}(\lambda,K_2-K_1)\mid K_1\rangle\mid^2
\end{eqnarray}
with ${\cal M}(\lambda,\nu)$ the intrinsic spherical component.
For dipole transitions between pear-shaped parity-doublet bands
${\cal M}(E1,0)=[3/(4\pi)]^{1/2} {\hat D}_z$, where the dipole
moment ${\hat {\bf D}}= e(N\sum_p {\bf r}_p-Z\sum_n {\bf r}_n)/A$.

In the strong coupling limit with two octupole minima at $\pm
\beta^{eq}_{30}$, the transition matrix element $D^t$ is
calculated as the intrinsic dipole moment at this deformation of
equilibrium. In this case, $\beta^{eq}_{30}=\beta_{30}^{\pi-}$,
the expectation value of $\beta_{30}$ in the first excited state
of negative parity, nearly degenerate with the g.s.

For shallow minima, closer to oscillation scenario, the degeneracy
between parity-doublet bands is shifted. The transition matrix
element in the intrinsic frame between the g.s. and the lowest
excited state of negative parity $\mid \pi -\rangle$ may be
approximated by a diagonal matrix element of the transition
operator in the mean-field state with the deformations
$\beta^{tr}_{30}$ fixed as the most probable by the transition
density. For a harmonic lowest-lying phonon one has the relation
$\beta_{30}^{tr}=0.63 \beta_{30}^{\pi-}$ which follows from the
one-dimensional harmonic oscillator.

We calculated expectation values of the electric dipole moment in
the intrinsic states as a sum of the macroscopic and
shell-correction parts, see e.g. \cite{Lea86,But91}. The
macroscopic part, derived within the Droplet Model in
\cite{Dor86}, was calculated as in \cite{Ska94}, i.e. without
assuming small deformations $\beta_{\lambda 0}$. It turns out
(Table \ref{tab:D0}) that intrinsic dipole moments at equilibrium
deformations for two configurations: $K=1/2$ and 3/2 (levels no 43
and 44) interpolate the values for $^{146,148}$Nd. Both
macroscopic and shell correction contributions have the same sign,
so no reduction of $D_0$ due to their cancellation is possible.
Thus, the dipole moments inferred from experiment do not fit the
picture of parity doublet bands.

\section{Summary}

The advanced time-delayed $\beta\gamma\gamma $(t) method has been
used to measure half-lives of 8 excited states in $^{147}$Nd.
Reduced transition probabilities were obtained for 30 transitions.
Twenty-four new $\gamma$-lines and 3 new levels have been
introduced into the decay scheme of $^{147}$Nd based on the
results of the $\beta$-gated $\gamma \gamma$ coincidence
measurement.

The potential energy surfaces on the ($\beta_{2}$,$\beta_{3}$)
plane and theoretical $|D_0|$ values suggest the presence of
octupole deformation in $^{147}$Nd at low excitation energies for
two configurations with K=1/2 and K=5/2. This suggestion is
supported by high experimental $|D_0|$ values in two even-even
neighbors of $^{147}$Nd. For the K=1/2 configuration we were able
to determine only lower limit of the dipole moment,
$|D_0|\geq$0.02 e$\cdot fm$. However non-observation of the K=5/2
parity doublet band may suggest that in $^{147}$Nd the $|D_0|$
values are as low as observed in other odd N isotopes at low
excitation energies. Probably strong octupole correlations should
be searched at higher excitation energies in the odd N isotopes
from the lanthanides octupole region.

\begin{acknowledgments}

One of us (E.R.) would like to thank the OSIRIS group for their
generous hospitality and for financial support during her stay in
Studsvik. This work was partially supported by Narodowe Centrum
Nauki under grant no. 2011/01/B/ST2/05131. M.K. and J.S. were
co-financed by LEA COPIGAL funds.

\end{acknowledgments}

\end{document}